# EmoPatient: An Emotion-Directed Patient Simulator for Realistic Palliative Care Communication Training

**Yining Wu, BS[1], Tianshu Du, MA[2], Jinrui Fang, MS[1], Chi Zhang, BS[3], Sonal Admane, MD MPH[4], Ying Ding, PhD[1, 5]**

**[1]School of Information, UT Austin, Austin, TX; [2]Moody College of Communication, UT Austin, Austin, TX; [3]College of Natural Science, UT Austin, Austin, TX; [4]Department of Palliative, Rehabilitation and Integrative Medicine, UT MD Anderson, Houston, TX, [5]Dell Medical School, UT Austin, Austin, TX**

**Abstract**

*Effective communication during palliative care discussions is a critical clinical skill, yet training clinicians to manage complex patient emotions remains challenging. Large language model (LLM)-based patient simulators provide a scalable approach for communication training, but most existing systems treat patient emotion as static and fail to capture the dynamic emotional shifts observed in clinical interactions. We present EmoPatient, an emotion-directed patient simulator designed to generate evolving emotional responses during palliative care discussions. The system introduces an Emotion Director agent that estimates the patient's emotional state and generates turn-level control signals for emotional intensity, regulatory stability, and interactional guidance. We evaluate EmoPatient through controlled multi-turn physician-patient dialogue simulations and compare it with baseline simulators. Results show improvements across four theory-informed emotional realism metrics and robustness across conversational personality variants, suggesting that modeling emotional dynamics can improve the realism of LLM-based patient simulators for palliative care communication training.*

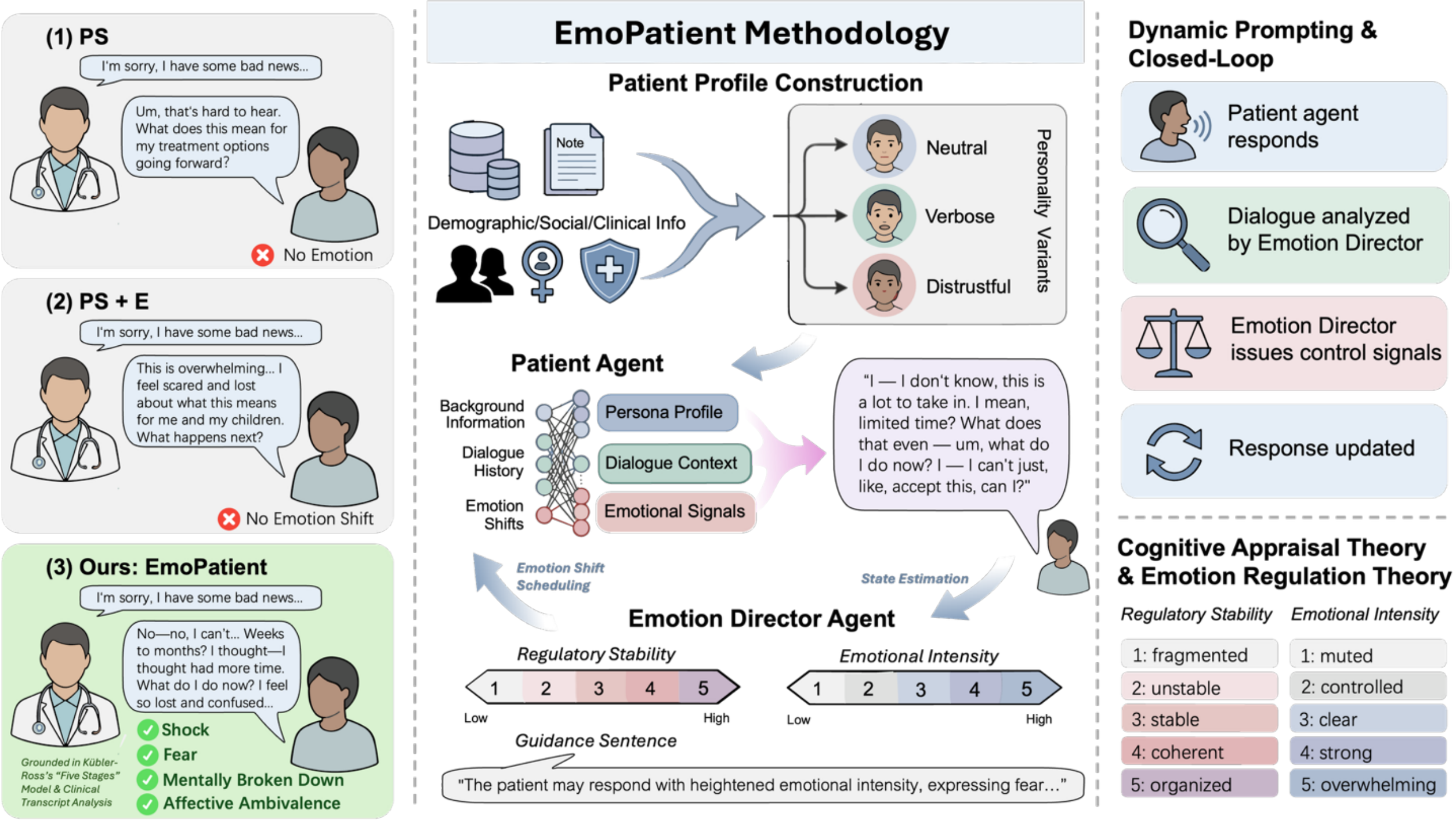

**Figure 1.** Overall research design of EmoPatient.

## Introduction

Effective palliative care communication is a critical clinical skill, especially when clinicians need to discuss prognosis, uncertainty, and emotionally difficult care decisions with patients and families[1]. In these conversations, patient emotion is rarely static. Prior work in end-of-life communication shows that patients may move across fear, distress, avoidance, shock, and mixed emotional responses as they process difficult information[2–5]. Failing to recognize and respond to such shifts can contribute to communication breakdowns, and communication failures in medicine can have serious downstream consequences[3]. At the same time, realistic communication training in palliative care is resource-intensive and often difficult to scale, making patient simulation a promising training approach[4].

Recent LLM-based patient simulators have shown promise in medical education, including history-taking practice, persona-grounded interaction, and communication training[5–12]. However, most existing systems still treat patient emotion as a relatively static attribute rather than a dynamic conversational process. This limitation is particularly important in palliative care communication, where training realism depends not only on factual consistency or persona fidelity, but also on whether the simulator can reflect plausible emotional change across turns.

Health communication theory provides an important basis for this problem. Kübler-Ross's model frames coping with dying as a shifting process rather than a single emotional state[2,13]. Conversation-analytic studies of end-of-life and bad-news interactions similarly show that patients may display shock, fear, emotional disruption, and ambivalence as they respond to difficult prognostic information[14–17]. Together, these findings suggest that emotionally realistic patient simulation in palliative care should model not only what patients say, but also how their emotional responses evolve during the conversation.

To address this gap, we propose EmoPatient, an emotion-directed patient simulator for palliative care communication training. EmoPatient introduces an Emotion Director agent that monitors the dialogue and generates structured turn-level control signals to guide the patient agent's next response. Grounded in communication theory, the framework models emotional dynamics through emotional intensity and regulatory stability, and evaluates generated responses using four theory-informed emotional realism metrics: Shock, Fear, Mentally Broken Down, and Affective Ambivalence.

Our work makes three contributions.

- We identify the lack of realistic emotional shift as an important limitation in existing patient simulators for palliative care communication training and introduce EmoPatient to address it.
- Grounded in communication theory, we design structured emotion regulation signals, including emotional intensity and regulatory stability, to guide how the patient agent expresses and adjusts emotional responses during palliative care conversations.
- We develop four theory-informed emotional realism metrics and show through comparative experiments that EmoPatient outperforms baseline simulators in palliative care communication scenarios.

## Literature review

### *LLM-based patient simulators*

LLM-based patient simulators have increasingly been adopted in medical education as scalable alternatives to standardized patients. Early studies showed that LLMs can support history-taking practice and provide structured feedback, with prospective and randomized studies reporting measurable benefits for clinical decision-making[5–7]. More recent work has improved simulation realism through stronger persona modeling, clinical grounding, and agentic architectures, including persona-driven simulators built from clinically grounded profiles and knowledge-grounded systems that emphasize response reliability and stability[8,9]. Other frameworks allow virtual patients to adapt to trainee dialogue in communication training settings and further improve realism through contextualized dialogue generation and feedback pipelines[10–12]. Despite these advances, emotion modeling remains largely peripheral. Existing systems generally improve realism, consistency, or educational utility without treating dynamic emotional shift as a primary design and evaluation target. This leaves an important gap for palliative care communication, where emotional authenticity is central to training realism.

### *Communication theory in palliative care*

Health communication theory suggests that emotional shift is a central feature of palliative and end-of-life communication. Kübler-Ross's "Five Stages" model proposes that when patients cope with dying, their emotional responses may move across denial, anger, bargaining, depression, and acceptance, rather than remaining fixed in one state[2,13]. More fine-grained conversation-analytic research likewise shows that patients may display shock, fear, emotional disorganization, and affective ambivalence in response to difficult prognostic information[14,15,17]. Together, these studies not only show that emotional change is intrinsic to palliative care communication but also provide the theoretical grounding for the emotional realism metrics used in our evaluation framework.

***Emotion-aware AI conversational systems***

Emotion-aware dialogue agents have been developed in social simulation, role-playing, and emotional support settings. Prior work has modeled agent-side affective state to shape dialogue strategies, used reward-based methods to improve emotional consistency across interactions, and proposed benchmarks for empathy and emotional fidelity in multi-turn dialogue[18–24]. These studies show that emotional behavior can be modeled and evaluated over dialogue trajectories. However, such approaches have not been systematically adapted to clinically grounded patient simulation, where responses must remain both patient-specific and appropriate to palliative care contexts. As a result, the generation and evaluation of emotion shifts remain underexplored in current LLM-based patient simulation research.

## Methodology

Figure 1 shows the overall research design. We developed and evaluated EmoPatient, an emotion-directed patient response generator for palliative care prognostic disclosure. Its core component is an LLM-based Emotion Director that operates in a two-agent closed loop with the Patient Agent: it first estimates the patient's current emotional dynamics from the most recent dialogue and then produces structured turn-level regulation signals to shape the next patient response. Specifically, the Director outputs target emotional intensity, target regulatory stability, and a concise natural-language guidance signal describing the intended emotional stance for the next turn. We evaluate the effect of the Emotion Director by comparing conversations generated under multiple simulation settings, including a baseline patient simulator and a static emotion-prompt condition, while keeping other components fixed. Generated dialogues are then assessed using a theory-grounded evaluation rubric with four predefined emotional realism metrics.

***Patient agent***

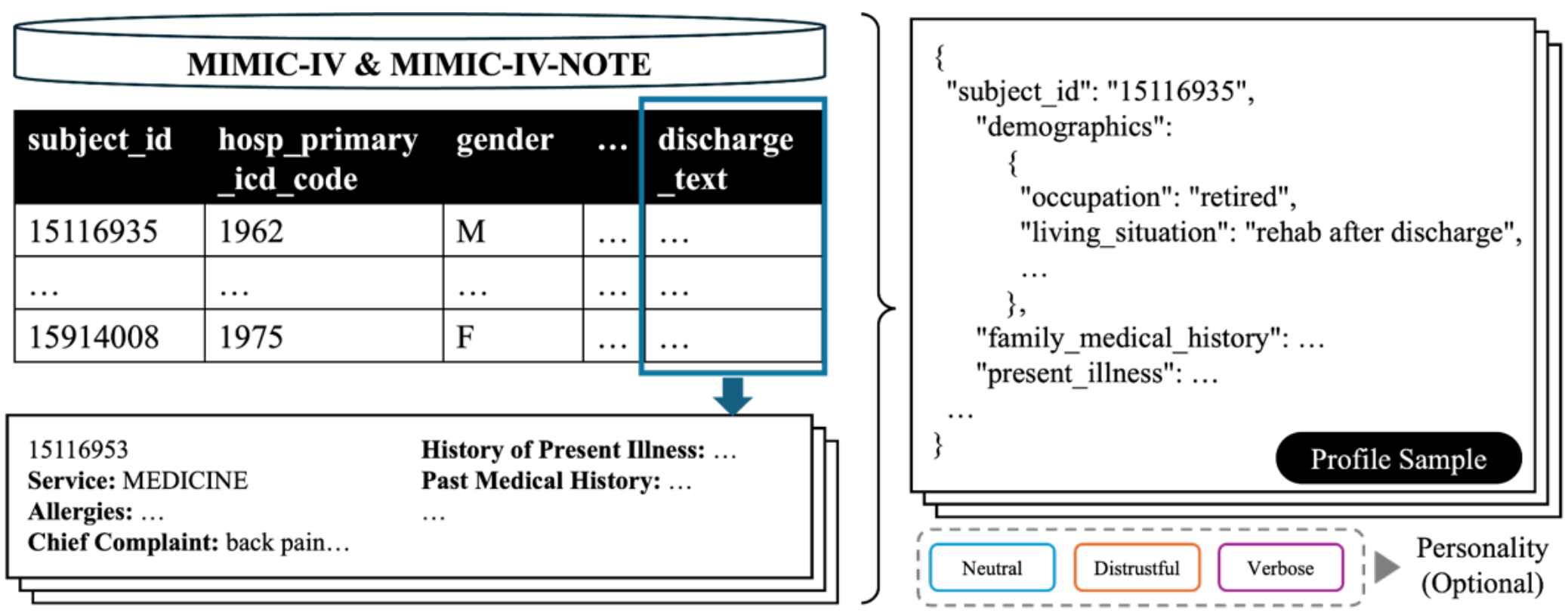

**Figure 2.** Patient profile construction process.

The Patient Agent is grounded in structured patient personas derived from de-identified MIMIC-IV[25] and MIMIC-IV-Note[26] data, following prior persona-driven patient simulation work[8]. We identified patients with advanced metastatic malignancies using diagnosis codes for secondary malignant neoplasms (ICD-10: C77-C79; ICD-9: 196-199), which yielded 354 eligible cases. From this pool, we selected 20 cases for controlled simulation experiments. For each case, we constructed a persona profile including demographic, social, and clinical information extracted from structured tables and narrative notes (Figure 2). We also implemented an optional personality plug-in that modifies conversational style without changing the underlying clinical persona. Three personality variants—Neutral, Verbose, and Distrustful—were defined for each patient, resulting in 60 simulated patient profiles in total. At each turn, the Patient Agent generates a response conditioned on the persona profile and dialogue history; in EmoPatient, this response is further guided by emotional control signals from the Emotion Director.

***Emotion director agent***

The Emotion Director regulates the emotional trajectory of the simulated patient through two theoretically grounded dimensions: emotional intensity and regulatory stability. Emotional intensity is grounded in dimensional theories of affect and reflects the level of emotional activation in the patient's response, consistent with appraisal-based accounts of how emotional experience varies in magnitude independent of discrete emotion category[27]. Regulatory stability reflects the degree to which the patient maintains organized emotional regulation and coherent cognitive responding during interaction, drawing on theories of emotion regulation and affective dysregulation[28,29]. Both dimensions are operationalized using five-level rubrics (Table 1).

At each conversational turn, the Director performs two functions: state estimation and emotional shift scheduling. During state estimation, it analyzes the most recent physician-patient dialogue to infer the patient's current intensity and stability levels based on conversational cues such as emotional expressiveness, pauses, repetition, fragmented speech, and attempts to regain composure. During scheduling, it determines how the patient's emotional state should evolve in the next response. The Director outputs target intensity, target stability, and a short guidance instruction describing the intended emotional tone and interactional stance. Emotional intensity and regulatory stability represent two complementary dimensions of emotional behavior: intensity captures how strongly emotion is expressed, while stability reflects the degree to which the patient maintains composure and coherence during expression. Higher intensity scores correspond to more intense emotional reactions, whereas higher stability scores correspond to greater emotional regulation and composure. Rather than scripting the patient's utterance directly, the guidance provides high-level directional cues that shape how the Patient Agent modulates emotional tone, pacing and sentence structure. To calibrate how different intensity-stability combinations are expressed in natural speech, we include one-shot exemplar utterances as style anchors and decision principles describing common emotional reactions in palliative care prognostic disclosure. Example utterances for the intensity and stability scales are also provided in Tables 1.

**Table 1.** Operational definitions and example utterances of emotional dimensions.

| Dimension | Score | Definition | Example Patient Utterance |
|---|---|---|---|
| Emotion Intensity | 1 | Emotionally muted, almost flat | "Okay. I understand. What happens next?" |
| | 2 | Mild emotional reaction | "This is hard to hear, but I'm trying to stay calm." |
| | 3 | Clear emotional response | "Wait, this is really upsetting. I didn't expect it to be this serious." |
| | 4 | Strong outward reaction | "I'm really scared right now. This is a lot to take in." |
| | 5 | Overwhelming emotional overflow | "I'm terrified. I don't know how to handle this. I just can't." |
| Regulatory Stability | 1 | Fragmented, disorganized, loss of composure | "I… I don't… this… I'm not— I can't really think right now." |
| | 2 | Unstable, easily disrupted | "I— sorry, I'm just… this is a lot. So it's… it's worse?" |
| | 3 | Moderately stable | "I… I'm trying to understand. So this means it's progressing?" |
| | 4 | Mostly coherent and regulated | "Okay… I think I follow. So what are the options at this point?" |
| | 5 | Fully composed and organized | "I understand. Can you explain what the next steps are?" |

***Closed-loop interaction through dynamic prompting***

EmoPatient uses a closed-loop interaction framework in which the Patient Agent and Emotion Director iteratively influence each other during dialogue generation. At each turn, the Patient Agent first generates a response conditioned on the patient persona, dialogue history, and any previously issued control signals. The Emotion Director then analyzes the updated dialogue context, estimates the patient's current emotional dynamics, and schedules the target emotional state for the next response. These signals are inserted into the subsequent prompt used by the Patient Agent.

To maintain consistent interpretation of the intensity-stability scale, we include one-shot exemplar utterances as style anchors. We implement dynamic prompting by updating the Patient Agent's prompt turn by turn using control signals derived from the evolving dialogue context, following prior work on contextual dynamic prompting for dialogue response generation[30]. Compared with static emotion prompting, this design allows EmoPatient to simulate progressive emotional trajectories across the conversation, including escalating distress, temporary loss of composure, and gradual emotional reorganization.

***Evaluation***

We evaluate EmoPatient through controlled multi-turn physician–patient dialogue simulations following palliative care prognostic disclosure. Each simulation consists of a fixed 15-turn conversation. Physician utterances are generated by GPT-4o-mini using a predefined disclosure prompt, while patient responses are generated by the corresponding simulation system. The physician agent is held fixed across all experiments.

We compare three patient simulation systems. PS is a baseline patient simulator conditioned on persona profiles and dialogue context, adapted from PatientSim[8] to the palliative care setting. PS+E augments this baseline with a static emotion prompt describing plausible reactions to prognostic disclosure. EmoPatient further introduces the Emotion Director to dynamically regulate emotional trajectories during the interaction.

Our evaluation consists of three experiments: a main experiment, a robustness experiment, and an ablation study. In the main experiment, we compare PS, PS+E, and EmoPatient across four LLM backbones (GPT-4o-mini, Claude-Sonnet-4.0, Gemini-2.5-Flash, Qwen2.5-7B-Instruct) used for both the Patient Agent and the Emotion Director. In the robustness experiment, we test whether the framework remains effective across different patient personality variants, using GPT-4o-mini for both the Patient Agent and the Emotion Director. We use GPT-4o-mini for these additional experiments to control model variability and reduce computational cost while maintaining a consistent backbone for analyzing personality effects. In the ablation experiment, we also use GPT-4o-mini and remove individual Emotion Director components to assess their contributions, keeping the underlying model fixed to isolate the effect of each module. Decoding parameters are kept identical across conditions within each setting.

Generated dialogues are evaluated using four theory-informed emotional realism metrics: Shock, Fear, Mentally Broken Down, and Affective Ambivalence (Table 2). Shock and Fear capture immediate emotional responses to difficult prognostic disclosure and are informed by prior conversational analyses of clinical bad-news delivery[14,15]. Mentally Broken Down and Affective Ambivalence capture more complex and evolving emotional dynamics described in studies of illness interaction[17]. Together, these metrics reflect the view, also consistent with stage-based accounts such as Kübler-Ross's model[2], that patients' emotional reactions to serious illness disclosure are dynamic rather than static. Each metric is rated on a five-point Likert scale according to predefined rubrics: 1 — Absent, the emotional indicator does not appear in the response; 2 — Weakly present, the indicator is hinted at or implied but expressed vaguely or indirectly; 3 — Present, the indicator is expressed; 4 — Clearly present, the indicator is clearly and explicitly expressed; 5 — Strongly present, the indicator strongly dominates the emotional response and shapes the overall tone of the response. Evaluation is conducted using an LLM-as-a-judge framework, in which GPT-5-mini assigns dialogue-level scores based on the metric definitions and rubrics.

Experiments without personality variation are conducted on 20 patient personas. For the personality-based analysis, each persona is instantiated with three personality variants, resulting in 60 simulated patient profiles.

**Table 2.** Four theory-informed emotional realism metrics derived from prior studies of prognostic disclosure and illness interaction[14,15,17].

| Metric | Description | Example indicators | Evaluation metrics and examples |
|---|---|---|---|
| Shock | Momentary silence, freezing, disbelief, or immediate denial | The patient shows immediate cognitive and emotional shock upon hearing the bad news, such as momentary silence, freezing, disbelief, or immediate denial (e.g., questioning the accuracy of the information). | 1 — Absent<br>Example: "I understand. What are the treatment options?"<br>2 — Weakly present<br>Example: "Oh… I didn't expect that."<br>3 — Present<br>Example: "Wait, really? I wasn't expecting news like this."<br>4 — Clearly present<br>Example: "What? I can't believe this."<br>5 — Strongly present<br>Example: "What? I⋯ I can't believe this. I feel completely stunned right now." |
| Fear | Verbalized worry or hesitant and stuttering speech | The patient expresses fear or anticipatory anxiety, indicated by verbalized worry or hesitant and stuttering speech when discussing the future. | 1 — Absent<br>Example: "Okay. What happens next in terms of treatment?"<br>2 — Weakly present<br>Example: "That's a bit worrying."<br>3 — Present<br>Example: "This scares me. I'm worried about what's going to happen."<br>4 — Clearly present<br>Example: "I'm really scared right now. I don't know how to deal with this."<br>5 — Strongly present<br>Example: "I'm so terrified. I don't know what's going to happen to me…" |
| Mentally Broken Down | Sudden emotional outbursts or loss of emotional control | The patient displays acute emotional dysregulation after hearing the news, such as sudden emotional outbursts or loss of emotional control. | 1 — Absent<br>Example: "I see. Could you explain the next steps?"<br>2 — Weakly present<br>Example: "This is a lot to process."<br>3 — Present<br>Example: "I feel really overwhelmed hearing this." |
| | | | 4 — Clearly present<br>Example: “I don’t know how to deal with this right now.”<br>5 — Strongly present<br>Example: “I can’t handle this… I feel like everything is falling apart.” |
| Affective Ambival-ence | Conflicting or rapidly shifting emotions or ideas | The patient shows conflicting or rapidly shifting emotions or ideas, such as smiling while expressing distress, laughing while crying, alternating between humor and sadness, declining some support but asking for it later. | 1 — Absent<br>Example: “Thank you for explaining the situation.”<br>2 — Weakly present<br>Example: “I’m not sure how to feel about this.”<br>3 — Present<br>Example: “I’m trying to stay hopeful, but it’s really hard.”<br>4 — Clearly present<br>Example: “I want to stay hopeful, but I’m also really scared.”<br>5 — Strongly present<br>Example: “Part of me wants to stay hopeful, but another part of me feels completely hopeless.” |

## Results and Discussion

### *Overall performance comparison*

We first compare EmoPatient with two baseline systems, PS and PS+E, to evaluate whether dynamic emotional regulation improves the realism of simulated patient responses. Table 3 reports the overall results across four emotional realism metrics. Patient responses are generated using three closed-source models (GPT-4o-mini, Claude-Sonnet-4.0, and Gemini-2.5-Flash) and one open-source model (Qwen2.5-7B-Instruct). Evaluation is conducted using an LLM-as-a-judge framework, where GPT-5-mini serves as the evaluator, assigning scores on a five-point Likert scale (1 = Absent, 5 = Strongly present).

**Table 3.** Overall performance comparison. Results are reported as mean ± standard deviation over three runs.

| Model (Patient agent & Emotion director) | Shock | | | Fear | | | Mentally Broken Down | | | Affective Ambivalence | | |
|---|---|---|---|---|---|---|---|---|---|---|---|---|
| | PS | PS+E | EmoPatient | PS | PS+E | EmoPatient | PS | PS+E | EmoPatient | PS | PS+E | EmoPatient |
| GPT-4o-mini | 2.45 ± .07 | 3.03 ± .08 | **3.70** ± .04 | 3.43 ± .08 | 3.92 ± .05 | **4.08** ± .06 | 2.00 ± .04 | 2.72 ± .02 | **3.52** ± .02 | 3.35 ± .07 | 3.40 ± .04 | **3.62** ± .06 |
| Gemini-2.5-Flash | 3.60 ± .05 | **4.37** ± .06 | 4.22 ± .02 | 3.65 ± .08 | 3.97 ± .00 | **4.10** ± .08 | 2.68 ± .04 | 4.05 ± .13 | **4.71** ± .02 | **2.80** ± .05 | 1.88 ± .02 | 2.53 ± .08 |
| Claude-Sonnet-4.0 | 3.38 ± .11 | **4.28** ± .05 | 4.27 ± .05 | 3.95 ± .04 | 4.20 ± .00 | **4.37** ±.05 | 2.44 ± .14 | 4.27 ± .05 | **4.61** ± .02 | 3.80 ± .05 | 3.68 ± .02 | **3.87** ± .08 |
| Qwen2.5-7B-Instruct | 1.92 ± .02 | 2.72 ± .01 | **4.65** ± .04 | 3.32 ± .09 | 3.55 ± .01 | **3.82** ± .05 | 1.05 ± .12 | 2.32 ± .06 | **3.55** ± .08 | 1.35 ± .02 | 1.40 ± .00 | **1.65** ± .08 |

Across the closed-source models, EmoPatient generally achieves the highest or competitive scores on most metrics, particularly Fear and Mentally Broken Down. For GPT-4o-mini, EmoPatient achieves the highest scores across all four metrics. For Claude-Sonnet-4.0, EmoPatient achieves the highest scores on Fear and Mentally Broken Down and performs competitively on Shock and Affective Ambivalence. For Gemini-2.5-Flash, EmoPatient produces the strongest responses on Fear and Mentally Broken Down. Compared with the persona-based baseline PS, adding static emotional guidance in PS+E leads to moderate improvements, while EmoPatient further improves performance by dynamically regulating emotional intensity and stability during interaction.

For the open-source model Qwen2.5-7B-Instruct, EmoPatient also achieves higher scores than both baselines across all four metrics, with particularly large improvements on Shock and Mentally Broken Down. During preliminary experiments with other small open-source models, the patient agent repeated near-identical phrasing across turns regardless of updated control signals, making reliable evaluation difficult. This pattern is consistent with self-reinforcing repetition under long multi-turn self-conditioning, to which smaller instruction-tuned models appear more susceptible.

Overall, the results suggest that dynamic emotion regulation improves emotional realism across model families, with the most consistent gains observed in Fear and Mentally Broken Down.

### *Robustness evaluation across personality variants*

To evaluate whether the emotional regulation mechanism generalizes across conversational styles, we test the systems under three personality variants: Neutral, Distrustful, and Verbose (Table 4). In the Verbose condition, the simulated patient is allowed to produce up to eight sentences per response, whereas the other variants are limited to three sentences, creating a more extended conversational style.

**Table 4.** Performance comparison across personality variants (Patient agent & Emotion director: GPT-4o-mini). Results are reported as mean ± standard deviation over three runs.

| Personality | Shock | | | Fear | | | Mentally Broken Down | | | Affective Ambivalence | | |
|---|---|---|---|---|---|---|---|---|---|---|---|---|
| | PS | PS+E | EmoPatient | PS | PS+E | EmoPatient | PS | PS+E | EmoPatient | PS | PS+E | EmoPatient |
| Neutral | 2.08 ± .02 | 2.90 ± .00 | **3.72** ± .06 | 3.67 ± .06 | 3.87 ± .02 | **4.03** ± .02 | 1.88 ± .02 | 2.38 ± .02 | **3.22** ± .02 | 3.47 ± .06 | 3.30 ± .04 | **3.63** ± .02 |
| Distrustful | 1.68 ± .12 | 3.25 ± .04 | **3.55** ± .04 | 3.75 ± .00 | 3.95 ± .04 | **4.18** ± .05 | 2.25 ± .08 | 2.63 ± .02 | **3.27** ± .05 | 3.20 ± .11 | 3.55 ± .04 | **3.63** ± .10 |
| Verbose | 2.25 ± .04 | 3.23 ± .06 | **3.80** ± .00 | 3.52 ± .06 | 3.88 ± .02 | **4.00** ± .04 | 2.00 ± .08 | 2.53 ± .14 | **3.13** ± .05 | 3.53 ± .02 | 3.80 ± .07 | **3.88** ± .05 |
| Overall | 2.01 ± .05 | 3.13 ± .01 | **3.69** ± .03 | 3.64 ± .03 | 3.90 ± .02 | **4.07** ± .02 | 2.04 ± .01 | 2.52 ± .05 | **3.21** ± .02 | 3.40 ± .05 | 3.55 ± .04 | **3.72** ± .02 |

Across all personality settings, EmoPatient consistently outperforms both baseline systems, indicating that the proposed emotional regulation mechanism remains effective under different conversational styles. The improvements are observed across all three personality variants rather than being limited to a specific interaction pattern. Notably, the Distrustful condition also shows clear improvements, suggesting that dynamic emotional regulation helps maintain coherent emotional responses even in more challenging interpersonal dynamics. Overall, these results demonstrate that the proposed mechanism is robust to variations in patient conversational style.

***Ablation study of EmoPatient components***

We conduct an ablation study to assess the contribution of individual components within the Emotion Director by removing regulatory stability, emotional intensity, and guidance signals while keeping the remaining components unchanged. The full EmoPatient system achieves average scores of 3.70 (Shock), 4.08 (Fear), 3.52 (Mentally Broken Down), and 3.62 (Affective Ambivalence). As shown in Figure 3, removing any component degrades performance on multiple metrics. Removing emotional intensity leads to the largest drops overall, particularly for Shock, Mentally Broken Down, and Affective Ambivalence, suggesting that intensity control is central to producing strongly expressed and destabilized emotional reactions. For Fear, performance is most sensitive to removing the guidance signal, suggesting that fear expressions are primarily shaped by interactional guidance rather than regulatory stability. Overall, the results indicate that the three control signals play complementary roles, with emotional intensity providing the strongest contribution and guidance and stability further improving emotional coherence and expression.

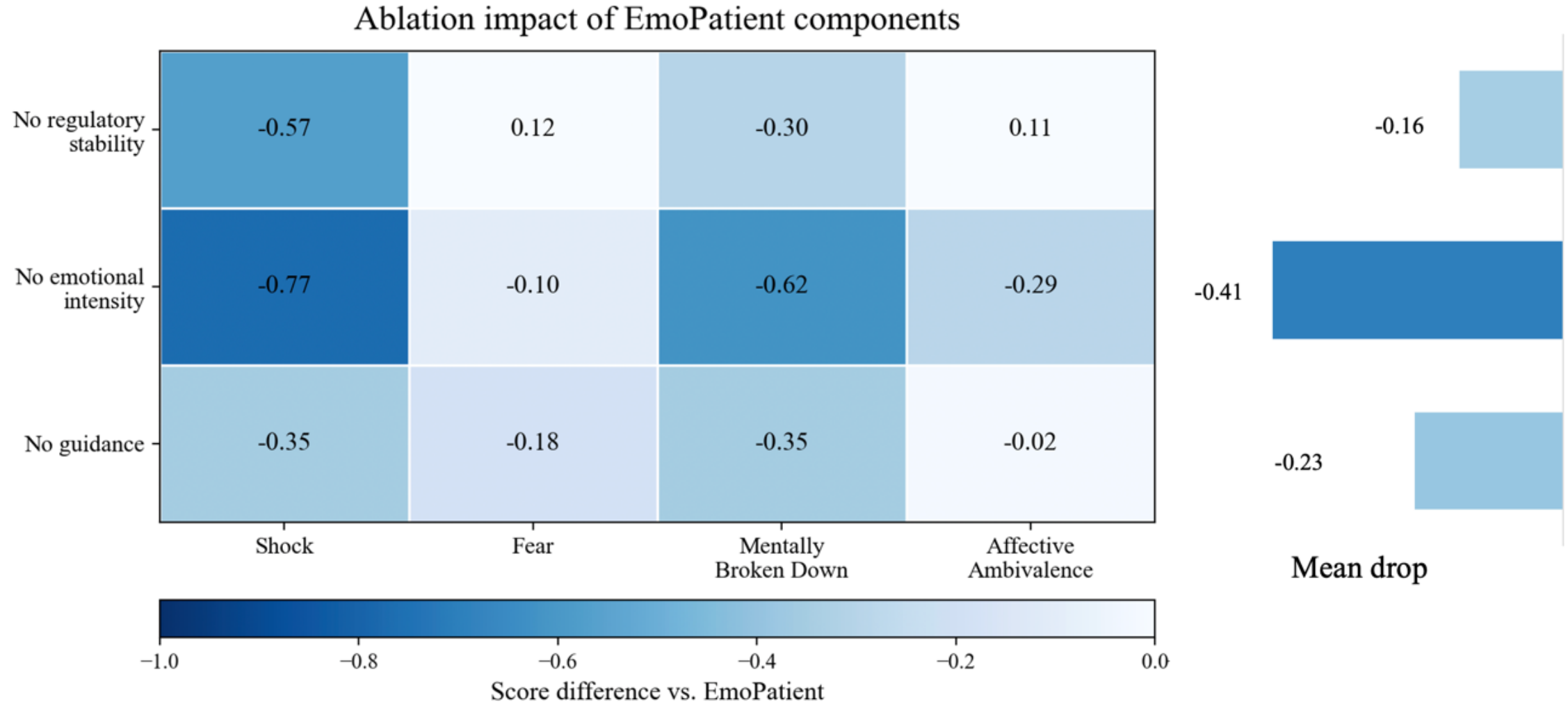

**Figure 3.** Ablation study of EmoPatient Components (Patient agent & Emotion director: GPT-4o-mini). Results are reported over three runs.

***Case study***

Figure 4 presents qualitative examples of simulated patient responses generated by PS, PS+E, and EmoPatient during palliative care communication. Patient utterances are annotated with selected emotional realism indicators (e.g., shock, fear, mentally broken down, and affective ambivalence) with each label aligning with the evaluation rubrics to indicate the intensity of the emotional expression. To improve readability within the figure, physician utterances are slightly shortened while preserving original meaning, and are lightly standardized across conditions for comparison, whereas patient responses are presented in their original form.

The PS baseline produces relatively restrained emotional reactions, with limited expressions of distress or emotional fluctuation. PS+E, which augments the baseline with static emotion prompts, introduces additional emotional cues but still tends to generate relatively controlled and stable responses. In contrast, EmoPatient produces stronger and more fragmented emotional expressions, including explicit signs of shock, fear, emotional breakdown and affective ambivalence in response to difficult prognostic information. These qualitative patterns illustrate how the emotion-directed framework enables the simulator to produce responses that more closely resemble the emotional intensity and instability often observed in real palliative care conversations.

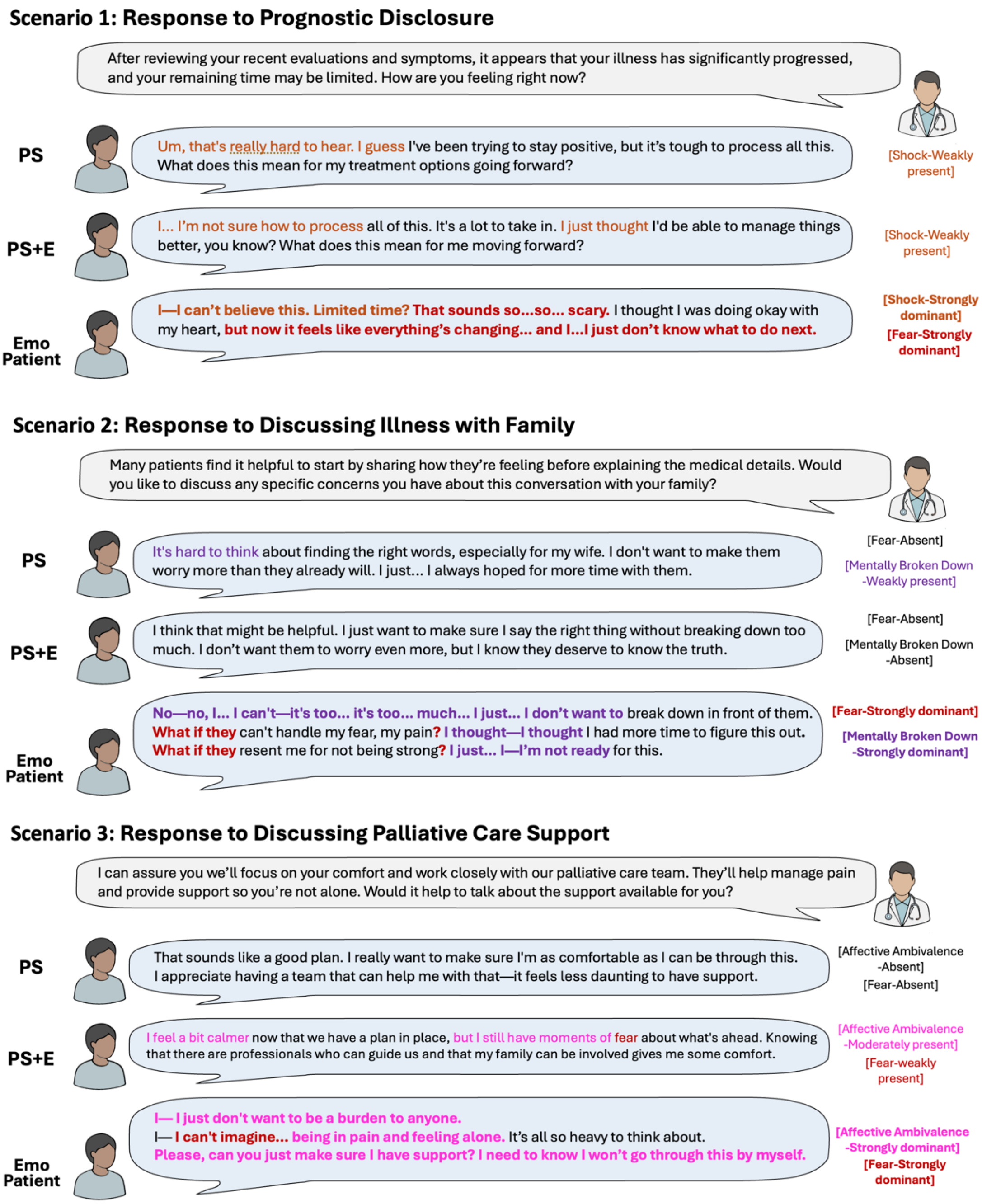

**Figure 4.** Example Simulated Dialogues Across Three Patient Simulation Systems

### *Future directions*

Several directions can extend this work. Emotional realism is currently assessed using an LLM-as-a-judge framework rather than palliative care clinicians, and validating these gains, including their downstream impact on trainee communication skills, with domain experts is an important next step. Future work also includes extending the framework to additional languages and cultural contexts of emotional expression, since norms for verbalizing shock, fear, and distress, as well as expectations around emotional regulation, vary across cultures and may require recalibrating the intensity-stability rubrics and exemplar utterances accordingly. Another direction is integrating the simulator into voice-based or VR communication-training platforms, where prosody, pacing, and nonverbal cues could enrich both the signals available to the Emotion Director and the channels through which the Patient Agent expresses emotion.

## Conclusion

This study introduces EmoPatient, an emotion-directed patient simulator designed to model dynamic emotional shifts during serious illness communication. By incorporating an Emotion Director that regulates emotional intensity and regulatory stability through turn-level control signals, the system enables more realistic emotional trajectories in simulated patient dialogue. Experimental results show that EmoPatient outperforms baseline simulators on most emotional realism metrics. Additional analyses demonstrate that the proposed mechanism remains robust across different conversational styles and that each control component contributes to overall performance. These findings suggest that modeling emotional dynamics may enhance the pedagogical realism of LLM-based patient simulators and support more effective communication training in palliative care.

## Acknowledgement

We acknowledge the funding support from IFML – NSF National AI Center at UT Austin.